# A Case Study in Survivable Network System Analysis

R. J. Ellison
R. C. Linger
T. Longstaff
N. R. Mead

*September 1998*

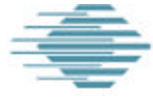

Carnegie Mellon
Software Engineering Institute

Pittsburgh, PA 15213-3890

# A Case Study in Survivable Network System Analysis

CMU/SEI-98-TR-014
ESC-TR-98-014

R. J. Ellison
R. C. Linger
T. Longstaff
N. R. Mead

*September 1998*

**Networked Systems Survivability Program**

**USAF Embedded Computer Resources Support Improvement Program (ESIP)**

This report was prepared for the

SEI Joint Program Office
HQ ESC/DIB
5 Eglin Street
Hanscom AFB, MA 01731-2116

The ideas and findings in this report should not be construed as an official DoD position. It is published in the interest of scientific and technical information exchange.

FOR THE COMMANDER

(signature on file)

Mario Moya, Maj, USAF
SEI Joint Program Office

This work is sponsored by the U.S. Department of Defense.
This report and the effort to produce it were sponsored by USAF Embedded Computer Resources Support Improvement Program (ESIP) and the U.S. Department of Defense.

# Table of Contents

# List of Figures

# List of Tables

# Abstract

This paper presents a method for analyzing the survivability of distributed network systems and an example of its application. Survivability is the capability of a system to fulfill its mission, in a timely manner, in the presence of attacks, failures, or accidents. Survivability requires capabilities for intrusion resistance, recognition, and recovery. The Survivable Network Analysis (SNA) method builds on the Information Security Evaluation previously developed by permitting assessment of survivability strategies at the architecture level. Steps in the SNA method include system mission and architecture definition, essential capability definition, compromisable capability definition, and survivability analysis of architectural softspots that are both essential and compromisable. Intrusion scenarios play a key role in the method. SNA results are summarized in a Survivability Map which links recommended survivability strategies for resistance, recognition, and recovery to the system architecture and requirements. This case study summarizes the application and results of applying the SNA method to a subsystem of a large-scale, distributed healthcare system. The study recommended specific modifications to the subsystem architecture to support survivability objectives. Positive client response to study recommendations suggests that the method can provide significant added value for ensuring survivability of system operations. As a result of this case study, the SNA method, artifacts, and lessons learned will be available to apply architectural analysis for survivability to proposed and legacy DoD distributed systems.

# 1 Network System Survivability

## 1.1 Survivability Concepts

As part of its Survivable Systems Initiative, the CERT® Coordination Center (CERT/CC) of the Software Engineering Institute (SEI) at Carnegie Mellon University is developing technologies and methods for analyzing and designing survivable network systems [Ellison 97, Linger 98, Lipson 97]. Survivability is defined as the capability of a system to fulfill its mission, in a timely manner, in the presence of attacks, failures, or accidents. Unlike traditional security measures that require central control and administration, survivability addresses highly distributed, unbounded network environments with no central control or unified security policy. Survivability focuses on delivery of essential services and preservation of essential assets, even when systems are penetrated and compromised. As an emerging discipline, survivability builds on existing disciplines, including security [Summers 97], fault tolerance [Mendiratta 92], and reliability [Musa 87], and introduces new concepts and principles.

The focus of survivability is on delivery of *essential services* and preservation of *essential assets* during attack and compromise, and timely recovery of full services and assets following attack. Essential services and assets are defined as those system capabilities that are critical to fulfilling mission objectives. Survivability depends on three key system capabilities: *resistance, recognition*, and *recovery*. Resistance is the capability of a system to repel attacks. Recognition is the capability to detect attacks as they occur, and to evaluate the extent of damage and compromise. Recovery, a hallmark of survivability, is the capability to maintain essential services and assets during attack, limit the extent of damage, and restore full services following attack.

---

® CERT is registered in the U.S. Patent and Trademark Office.

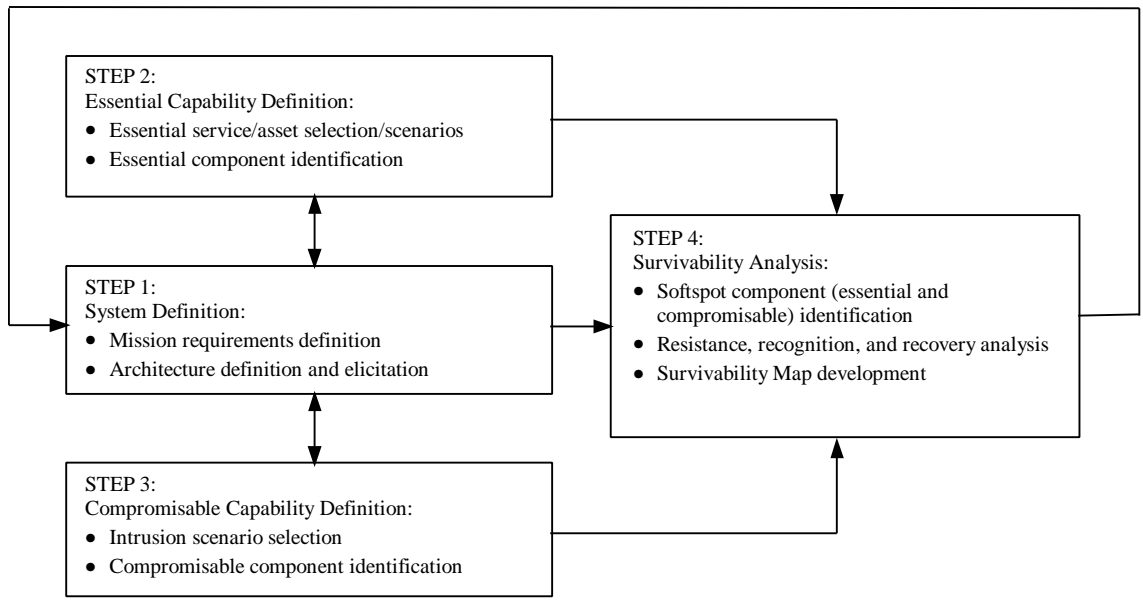

*Figure 1: The Survivable Network Analysis Method*

## 1.2 The Survivable Network Analysis Method

The Survivable Network Analysis (SNA) method for assessing and improving the survivability of network architectures is depicted in Figure 1. The method builds on the Information Security Evaluation method[1] by permitting the evaluation of a distributed architecture rather than focusing on the site-level security. The method can be applied to an existing or proposed system by a small team of trained evaluators through a structured interaction with system personnel composed of several meetings and working sessions.

The method is composed of four principal steps, as follows. In step 1, mission objectives and requirements for a current or candidate system are reviewed, and the structure and properties of its architecture are elicited. In step 2, essential services (services that must be maintained during attack) and essential assets (assets whose integrity, confidentiality, availability, and other properties must be maintained during attack) are identified, based on mission objectives and consequences of failure. Essential service and asset uses are characterized by *usage scenarios*. These scenarios are mapped onto the architecture as execution traces to identify corresponding *essential components* (components that must be available to deliver essential services and maintain essential assets). In step 3, *intrusion scenarios* are selected based on the system environment and assessment of risks and intruder capabilities. These scenarios are likewise mapped onto the architecture as execution traces to identify corresponding *compromisable components* (components that could be penetrated and damaged by intrusion). In step 4, *softspot components* of the architecture are identified as components that are both es-sential and compromisable, based on the results of steps 2 and 3. The softspot components and the supporting architecture are then analyzed for the three key survivability properties of resistance, recognition, and recovery. The analysis of the "three R's" is summarized in a *Survivability Map,* as shown in Figure 2. The map is a two-dimensional matrix that enumerates, for every intrusion scenario and corresponding softspot effects, the current and recommended *architecture strategies* for resistance, recognition, and recovery. The Survivability Map provides feedback to the original architecture and system requirements, and may result in an iterative process of survivability evaluation and improvement.

---

[1] Fraser, B.; Konda, S.; Lipson, H.; Longstaff, T.; & Alberts, C. *Information Security Evaluation: Site Coordinator's Guide* Pittsburgh, Pa.: Software Engineering Institute, Carnegie Mellon University, June 1998.

| **Intrusion Scenario** | **Resistance Strategy** | **Recognition Strategy** | **Recovery Strategy** |
|---|---|---|---|
| (Scenario 1) … | **Current:** | **Current:** | **Current:** |
| | **Recommended:** | **Recommended** | **Recommended:** |
| (Scenario n) | **Current:** | **Current:** | **Current:** |
| | **Recommended:** | **Recommended:** | **Recommended:** |

*Figure 2.   Survivability Map Template*

# 2 Sentinel: The Case Study Subsystem

Management of mental health treatment is often performed as a manual process based on hand-written forms and informal communication. Substantial time and effort are consumed in coordination of various treatment providers, including physicians, social service agencies, and healthcare facilities. CarnegieWorks, Inc. (CWI) is developing a large-scale, comprehensive management system to automate, systematize, and integrate multiple aspects of regional mental health care. The CWI system, named Vigilant, will ultimately be composed of some 22 subsystems operating on a distributed network of client and server computers, and will maintain a large and complex database of patient and provider records. A vital part of the Vigilant system is development and management of *treatment plans*. A treatment plan is developed for a *patient* by a *provider*. The *problems* of each patient are identified, together with a set of *goals* and *actions,* including medication and therapy, to achieve those goals. Each treatment plan is carried out by an interdisciplinary and interorganizational *action team* composed of providers. An *affiliation* is an organization that provides healthcare services, possibly to many patients. Treatment plan development and management and action team definition and coordination are key functions of the Sentinel subsystem. As a subsystem of Vigilant, Sentinel interacts with providers, affiliations, and other subsystems. It maintains the action teams and treatment plans as part of the Vigilant patient database, and applies regulatory and business rules for treatment plan development and validation. Because of the critical nature of mental health treatment, the need to conform to regulatory requirements, and the severe consequences of system failure, survivability of key Sentinel capabilities has been identified by CWI personnel as extremely important.

# 3 Applying the Survivable Network Analysis Method to Sentinel

## 3.1 Method Application

The SNA method was applied to the Sentinel subsystem through a structured series of meetings between the analysis team and project personnel (customer and development team), interleaved with analysis team working sessions, as shown in Figure 3.

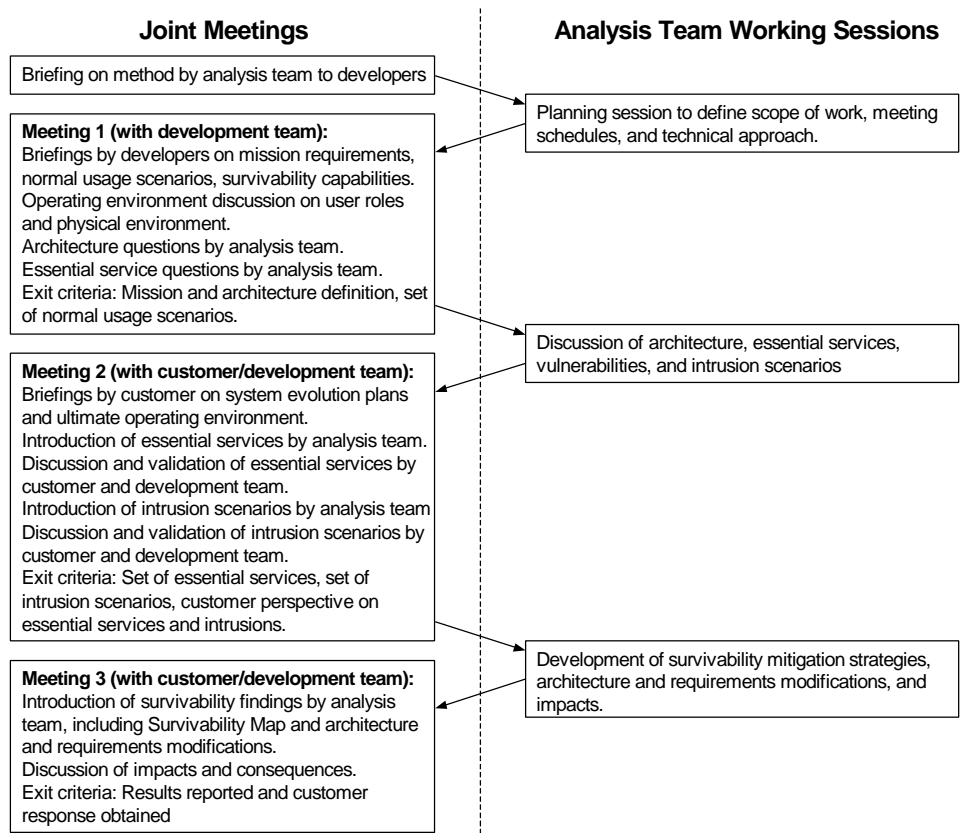

*Figure 3.    Meetings and Working Sessions in SNA Method Application*

The objective of the first meeting was to obtain as much information as possible about the subsystem and its mission and architecture. The development team briefed this material and provided supporting documents. The second meeting included the customer as well as the development team, and was used to understand the ultimate operating environment of the subsystem and future plans for the entire system. At this meeting, the analysis team validated the selection of essential services and assets for Sentinel, as well as the definition of system user types and characteristics. The analysis team also introduced and validated the set of intrusion scenarios to be applied to Sentinel. At the final meeting the analysis team presented its findings on Sentinel survivability. Proposed mitigation strategies in terms of resistance, recognition, and recovery were presented, and possible architectural modifications and requirements impacts were discussed. Customer reaction to the recommendations was positive. Between these meetings, the analysis team met in working sessions to assess the subsystem and its vulnerabilities, and to develop survivability recommendations. These meetings and working sessions carried out the SNA steps as described below.

### Step 1: System Definition

*Mission Requirements Definition*

The following normal usage scenarios (NUS) elicited from Sentinel requirements documentation characterize principal mission objectives of the subsystem. Each scenario includes a statement of the primary Sentinel responsibility with respect to the scenario:

- NUS1: Enter a new treatment plan. A provider assigned to a patient admitted into an affiliation performs an initial assessment and defines a treatment plan, specifying problems, goals, and actions. Sentinel must apply business rules to treatment plan definition and validation.

- NUS2: Update a treatment plan. A provider reviews a treatment plan, possibly adding or changing problems, goals, or actions, and possibly updating the status of these items. Sentinel must apply business rules to treatment plan update and validation.

- NUS3: View a treatment plan. A provider treating a patient views a treatment plan to learn the status of problems, goals, and actions. Sentinel must ensure that the plan displayed is current and valid.

- NUS4: Create or modify an action team. A provider defines or changes the membership of a treatment team in an affiliation for a patient. Sentinel must ensure that the treatment team definition is current and correct.

- NUS5: Report the current treatment plans in an affiliation. An administrator views the current state of her affiliation's treatment of a patient or set of patients. Sentinel must ensure that the treatment plan summaries are current and correct.

- NUS6: Change patient medication. A provider changes the medication protocol in a treatment plan for a patient, possibly in response to unforeseen complications or side effects. Sentinel must ensure that the treatment plan is current and valid.

*Architecture Definition and Elicitation*

The original Sentinel architecture obtained from design documentation is depicted in simplified form in Figure 4. Execution traces of the normal usage scenarios identified in step 1 were used by the evaluation team to illuminate and understand architectural properties. The traces revealed component sequencing within the architecture, as well as reference and update of database artifacts.

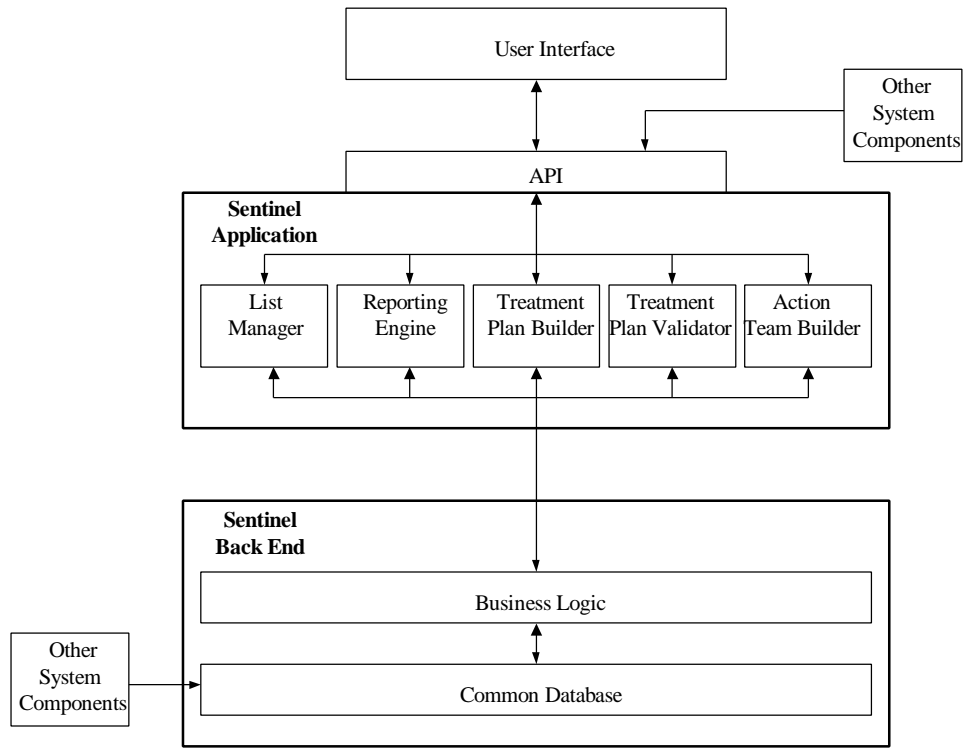

*Figure 4. Original Sentinel Architecture*

Architecture component functions are summarized as follows:

- User Interface: resides outside of Sentinel to allow a single User Interface to serve multiple subsystems and components.

- API: provides synchronous RPC and asynchronous messaging facilities for use by the User Interface and other system components.

- List Manager: maintains lists including patients, affiliations, providers, action teams, and relations among them.

- Reporting Engine: provides read-only viewing and reporting of Sentinel artifacts, including current treatment plans and their histories.

- Treatment Plan Builder: creates treatment plans for patients, including problems, goals, and actions.

- Treatment Plan Validator: checks the completeness and consistency of treatment plan development and modification.

- Action Team Builder: provides capability to define and modify action team membership.

- Business Logic: contains enterprise-defined business rules, including validation checks for treatment plan development and logging triggers that manage change control of sensitive data.
- Database: sentinel shares access to a common database with other subsystems and components.

**Step 2: Essential Capability Definition**

*Essential Service/Asset Selection/Scenarios*

Essential services and assets represent critical system capabilities that must survive and be available during intrusions. Criticality is based on analysis of mission objectives, risks and consequences of failure, and availability of alternatives. Such an analysis may result in selection of any number of essential services and assets, and may stratify them into survivability classes of varying criticality. The survivability analysis of the Sentinel subsystem was carried out together with CWI personnel, and was based on the normal usage scenarios identified in step 1. The analysis resulted in selection of a single essential service, namely, NUS3, the capability to view treatment plans. This service, more than any other, was deemed essential to delivery of mental health treatment because providers depend on real-time, on-demand access to treatment plans in clinical situations, particularly in cases of medication or therapeutic problems of an emergency or life-critical nature. The other normal usage scenarios could be postponed for hours or even days in the event of system intrusion and compromise. The analysis also identified a single essential asset, namely, the treatment plans themselves. Preservation of treatment plan integrity and confidentiality was deemed essential to meeting Sentinel mission objectives. The other Sentinel artifacts, such as action teams, affiliations, and providers, could all be reconstructed or updated hours or days after intrusion with no irreversible consequences.

*Essential Component Identification*

Essential system components are those components that participate in delivery of essential services and preservation of essential assets. The execution trace of the NUS3 scenario revealed the reporting engine and the database components, as well as their supporting components and artifacts, are essential to maintaining the capability to perform the scenario. As essential assets, the integrity and confidentiality of treatment plans depends on database components for security and validation.

**Step 3: Compromisable Capability Definition**

*Intrusion Scenario Selection*

Based on the system environment and assessment of intruder objectives and capabilities, the following five intrusion usage scenarios (IUS) were selected as representative of the types of attacks to which Sentinel could be subjected. Each scenario is preceded by an IUS number and type of attack (shown in parentheses), and followed by a brief explanation:

- IUS1 (Data Integrity and Spoofing Attack): An intruder swaps the patient identification of two validated treatment plans.

  Sentinel performs validation of treatment plans before entering them into the database. In this scenario, an intruder accesses the database server to corrupt treatment plans without using the Sentinel client, but rather by spoofing a legitimate client.

- IUS2 (Data Integrity and Insider Attack): An insider uses other legitimate database clients to modify or view treatment plans controlled by Sentinel.

  The database security assumes that clients have exclusive write access to specific database tables. While the IUS1 scenario attempts to access the database directly, this scenario examines inappropriate access through other database clients.

- IUS3 (Spoofing Attack): An unauthorized user employs Sentinel to modify or view treatment plans by spoofing a legitimate user.

  Some terminal access points for Sentinel are located in public areas, and hence are not as physically secure as those in private offices. This scenario illustrates opportunistic use of an unoccupied but logged-in terminal by an illegitimate user who spoofs the legitimate logged-in user.

- IUS4 (Data Integrity and Recovery Attack): An intruder corrupts major portions of the database, leading to loss of trust in validated treatment plans.

  Scenarios IUS1 and IUS2 assume a sophisticated attacker who targets and recognizes specific treatment plans, and modifies only a few fields. This scenario assumes a brute-force corruption of the database, leading to large-scale loss of trust and potential denial of service during massive recovery operations.

- IUS5 (Insider and Availability Attack): An intruder destroys or limits access to the Sentinel software so it cannot be used to retrieve treatment plans.

  This scenario could be as simple as removing the Sentinel software, or could involve attacks on the network or application ports to limit application access.

*Compromisable Component Identification*

Compromisable system components are those components that can be accessed and potentially damaged by intrusion scenarios. The execution traces of the five IUS scenarios revealed the following component vulnerabilities:

- IUS1: This scenario compromises the treatment plan component. There were no validity checks made on treatment plans after the initial entry.

- IUS2: This scenario compromises the treatment plan component. The treatment plan changes might be consistent but made by an improper agent.

- IUS3: This scenario compromises the treatment plan component. The majority of system users would object to logging into the system repeatedly as a way to continually monitor the validity of the user. The system had not considered those terminals which were in open areas easily accessible by unauthorized users.

- IUS4: This scenario compromises the treatment plan component. Database recovery required higher priority with respect to operations.

- IUS5: All software components of the Sentinel subsystem are affected by this scenario. While there were implicit user requirements on availability, it had not been considered in the architecture.

**Step 4: Survivability Analysis**

*Softspot Component Identification*

As noted earlier, softspot components are those components that are both essential and compromisable. The foregoing analysis shows that the (essential service) reporting engine component and the (essential asset) database treatment plan component can both be compromised in a variety of ways. The survivability analysis focuses on the essential services and assets that these components provide in fulfilling the mission objectives of the system.

*Resistance, Recognition, and Recovery Analysis*

Analysis of the three R's resulted in the Survivability Map depicted in Table 1(ID stands for identification, TP for treatment plan, UI for user interface, and DB for database). The recommendations in Table 1 are annotated with reference numbers {1} to {6} that correlate with changes to the architecture defined in Figure 5. Development of the table began by matching each intrusion scenario trace (created in step 3 above) to the softspot components. Each trace was first checked for all current resistance (protection) components in the architecture that would increase the difficulty experienced by an intruder in reaching the softspots referenced in the trace. Because no detailed implementation information was available to identify specific vulnerabilities in these resistance components, an assumption was made that any vulnerabilities in them would be found and corrected over time. The greater the resources available to an intruder, however, the less time a resistance component will be completely effective. The current resistance components are described in the resistance column of the Survivability Map for each scenario.

For the recognition column, a process similar to the resistance analysis was followed. To assess the effectiveness of current recognition components, a number of assumptions were made and listed in the Survivability Map. For example, in scenario IUS3 in Table 1, there is a documented assumption that a provider will become suspicious when there are a large number of denied accesses to treatment plans reported to some party. If this assumption is not valid, then there are no current recognition strategies associated with this scenario.

For the recovery column, assumptions were made regarding common database management facilities (standard backup and recovery of the database itself and version control of the Sentinel software). Table entries for current recovery strategies included these assumptions, so that if in fact they are not satisfied in the final system, the recovery strategy will be less effective than that described in the Survivability Map. However, the assumptions for the current recovery strategies take into account standard practice with regard to distributed database systems.

Once all of the current resistance, recognition, and recovery strategies were identified, gaps and weaknesses were analyzed for common points in the architecture where a particular survivability improvement could address multiple scenarios or multiple strategies. These high-leverage recommendations are listed in a consistent form and identified as a common recommendation. Other gaps identified by a lack of an existing strategy in any of the resistance, recognition, or recovery columns were also addressed. For the resistance column, recommendations were made even where an existing resistance mechanism existed, as this mechanism can be expected to degrade over time. Ultimately, it is up to the system architect to determine the cost-benefit of implementing these recommendations. The Survivability Map can help an architect determine the impact of accepting risks associated with weaknesses in the resistance, recognition, or recovery columns, as these are correlated to the intrusion scenarios that affect the essential services or assets of the system. In Table 1, a number of gaps and assumptions are identified in the current resistance, recognition, and recovery strategies. Of particular interest to an architect are those recommendations that deal with multiple intrusion scenarios. For example, adding a crypto-checksum to the validation of a treatment plan addresses several scenarios.

| Intrusion Scenario | Resistance Strategy | Recognition Strategy | Recovery Strategy |
| --- | --- | --- | --- |
| IUS1:<br><br>Intruder swaps the ID of two validated TPs. | **Current:**<br>Two passwords are required for TP access. | **Current:**<br>Logging of changes made to DB.<br>Provider may recognize an incorrect TP. | **Current:**<br>Built-in recovery in commercial DB.<br>Backup and recovery scheme defined. |
| | **Recommended:**<br>Implement strong authentication supported in a security API layer. {1} | **Recommended:**<br>Add crypto-checksum when TP is validated.{3} Verify crypto-checksum when TP is retrieved. {4} | **Recommended:**<br>Implement a recovery mode in the user interface to support searching for and recovering incorrect TPs. {1} |
| IUS2:<br><br>Outside agents exercise (legitimate) access to DB fields controlled by Sentinel. | **Current**:<br>Security model for DB field access. | **Current**:<br>None. | **Current**:<br>Scrap data and start over, or find an early backup and verify each entry. |
| | **Recommended**:<br>Need to verify the security model in light of module addition and integration. | **Recommended**:<br>Perform a validation on access of a TP for verification. {2} Add crypto-checksum when TP is validated.{3} Verify this checksum when TP is retrieved. {4} | **Recommended**:<br>Scan DB for invalid crypto-checksums and/or invalid TPs and recover to last known correct TP. {4} |
| IUS3:<br><br>An unauthorized user employs Sentinel to modify or view TPs by spoofing a legitimate user. | **Current**:<br>None. No timeout is specified so that anyone can use a logged in but vacated terminal. However, intruder only has access to logged in user's TPs | **Current**:<br>None, except for unusual number of denied accesses to TPs as an intruder attempts to locate particular TPs. | **Current**:<br>Can get list of modified TPs through the spoofed users transaction history. Manually recover each modified record. |
| | **Recommended**:<br>Add a short logout timeout for any terminals in uncontrolled areas (not physician's offices). {1} | **Recommended**:<br>Add logging, access control, and illegal access thresholds to the security API. {1} | **Recommended**:<br>Develop a recovery procedure and support it in the UI. {1} |
| IUS4:<br><br>Intruder corrupts DB leading to loss of trust in validated TPs. | **Current**:<br>Security model in the DB protects data against corruption. | **Current**:<br>None, except when provider happens to recognize a corrupted TP. | **Current**:<br>Locate an uncorrupted backup or reconstruct TPs from scratch. |
| | **Recommended**:<br>Implement live replicated DB systems that cross check for validity (supported in many commercial DB systems). {5} | **Recommended**:<br>Add and check crypto-checksums on records in the DB. {3} {4} | **Recommended**:<br>Reduce the backup cycle to quickly rebuild once a corrupted DB is detected. {5} |
| IUS5:<br><br>Intruder destroys the Sentinel software so it cannot be used to retrieve TPs | **Current**:<br>Keep originals available. | **Current**:<br>System doesn't work. | **Current**:<br>Reload the system from originals. |
| | **Recommended**:<br>Keep a spare CD available for quick recovery | **Recommended**:<br>None. Easy to detect this one. | **Recommended**:<br>Fast recovery from CD. Create a small sub-system that can retrieve TPs while Sentinel is down or being upgraded. {6} |

*Table 1.   Sentinel Subsystem Survivability Map*

The modified architecture resulting from the Survivability Map analysis is depicted in Figure 5, with additions and changes shown with dashed lines and shading. Many of the recommendations in the Survivability Map affect the same architectural component. To further illustrate the overlaps, reference number annotations {1} to {6} attached to the recommendations are included in the modified architecture. In this way, it was easy to determine which of the recommendations addressed multiple intrusion scenarios. With limited resources to mitigate these risks, this view of the recommendations can help the architect allocate resources to high-impact modifications of the architecture. As the modified architecture was formed to address the recommendations in the Survivability Map, several natural locations emerged in the existing architecture where implementation of the recommendations could be localized with minimal impact to the overall system. This was primarily due to the functional decomposition used in the original architecture. It is also likely that the evaluation of the scenarios led to the formation of recommendations that were natural to the architecture, since in executing the scenarios over the architecture, the impact on individual modules was evident.

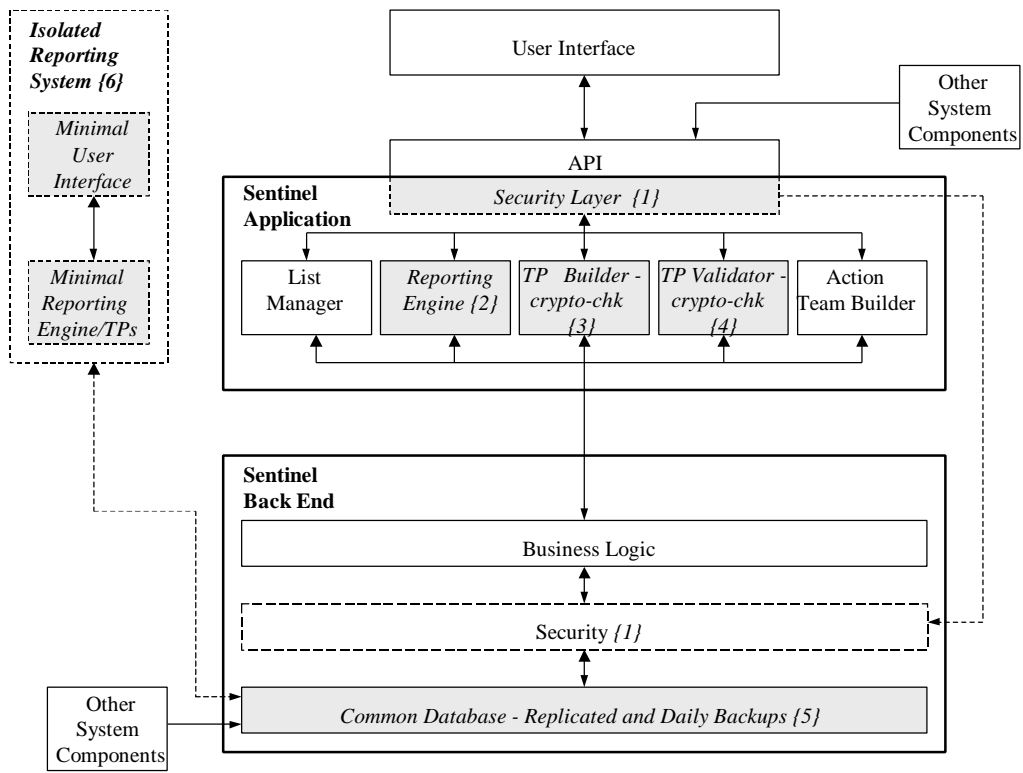

*Figure 5.   Sentinel Architecture with Survivability Modifications*

To support the essential service (treatment plan display) and asset (treatment plans) identified in the earlier stages of the process, a simple new component {6} was added outside the original architecture that could serve the purpose of retrieving treatment plans if the primary system should fail for any reason. With this external component, intrusion scenario IUS5 was addressed. This new component had minimal impact on the original architecture, since it identified a distinctly separate software program used to interface with the underlying database. It is possible, depending on the selection of the database system, that this small component could be included in the procurement of the software as a simple database retrieval program. In addition, to address the validation of treatment plans read from the database (not simply saved to the database), there was a sequencing recommendation ({2}, {3}, and {4}) that all data retrieved from the database would pass through the validation module to verify the correctness of the crypto-checksum.

A proactive validation function was also recommended, whereby the validation module would retrieve treatment plans from the database during idle time to continuously validate the data saved in the database against the saved crypto-checksums.

To address IUS3, it was desirable to add a security layer {1} to the architecture between the user interface and the other parts of the Sentinel architecture. This provided a location for monitoring and logging activity between the user interface and the Sentinel subsystem. This is especially important if the recommendations on the user interface (documented in IUS1, IUS3, and IUS4) were not implemented (these were out of scope for the Sentinel development team). The security layer provides functionality for passing user credentials to the database for access control in addition to providing intrusion detection, timeout information, and other security-relevant functions.

Several of the recommendations did not address the view of the architecture presented in Figure 5 directly, but were concerned with the use of the architecture. For example, the recommendation in IUS2 calling for the validation of the security model in the Sentinel back end system {5}. This is an example of an architectural requirement that is expressed in the Survivability Map, but is difficult to capture in the common "topology" view of the architecture. These recommendations are mapped to specific components in the topology view; however, the changes to these components are not evident in their implementation, but rather in the process of their implementation.

In addition to architectural analysis, these survivability findings can also be reflected in modifications to Sentinel requirements. The Mission Requirements Definition of step 1 revealed few specific survivability or security requirements for the Sentinel subsystem, other than requiring 1) validation of treatment plan data, 2) utilization of some security features built into the standard login process and the database, and 3) a development strategy that would permit easy modification to add security features. Changes are needed at the highest level to two areas of the requirements. Under survivability conditions, there is a critical need for providers to view treatment plans within a reasonable time.

In addition, there is a need to protect the integrity of the treatment plans in the database. These high-level requirements might be stated as follows:

- The treatment plan data shall be viewable within xx seconds of request under nominal conditions. The treatment plan data shall be viewable within yy seconds (minutes) of request during recovery.
- Resistance and recognition techniques shall be used to protect the integrity of the treatment plan data under intrusion scenarios IUS1 through IUS5.

These requirements can be refined to encompass software, procedural, and hardware requirements. The software requirements might be:

- An emergency reporting system shall allow treatment plans to be viewed during recovery.
- Treatment plans shall be validated when they are read and written. If a treatment plan is invalid, the last valid version of the treatment plan shall be recovered.
- Encrypted checksums shall be used to protect the integrity of the treatment plans.
- The selected database software shall support replication.

The procedural requirements might be:

- The Sentinel software shall be backed up on CD.
- Daily backups of the database shall be performed.

The hardware/operating system requirement might be:

- Workstations located in public areas shall have a short timeout based on inactivity. There shall be login access thresholds for incorrect logins.

# 4  Lessons Learned

The SNA method is under continuing development and additional case studies are planned. Lessons learned at this stage focus on the validity of the initial assumptions and objectives of the method, as well as refinements that can be explored in future case studies. The Sentinel case study began with three assumptions:

- Survivability strategies could be organized in terms of resistance, recognition, and recovery.
- The analysis should focus on early phases of the life cycle, specifically, on the mission requirements, as they represent the essential services and assets of the system, and on the architecture, as it represents the components that must be survivable and the strategies for achieving survivability.
- The application logic rather than the system infrastructure should bear a significant portion of the responsibility for implementation of survivability strategies [Saltzer 84].

The case study supported these assumptions. Organization of survivability strategies in terms of resistance, recognition, and recovery was straightforward and easily communicated to the customer. Identification of essential services and assets was a critical step in limiting the scope of the analysis, as well as in reducing the number and scope of architectural revisions which the customer should consider. The Sentinel subsystem examined in this study was just entering its implementation phase; future studies should include a need to reengineer existing systems.

The success of the SNA method depends on the effectiveness of the recommendations, that is, achievement of a modified system that is by some set of measures more survivable. Of equal importance is whether the customer can incorporate the recommendations into the existing software development process, and thus be able to adopt the suggested changes. Because the survivability recommendations for Sentinel concentrated on refining an existing architecture rather than requiring a redesign, they did satisfy this criterion.

While most of the recommendations focused on revisions to the application architecture, several suggested changes in design and implementation or in operations and procedures to support survivability in the existing architecture. The study did raise some issues of extensibility, that is, could the proposed architecture support the functionality desired in later versions from a survivability perspective. Analysis of extensibility could be an important aspect of future studies. The recommendations produced in this study were able to take advantage of existing system features to support reliability and fault tolerance, such as the transactional support and recovery mechanisms provided by the relational database.

While the study did not involve extensive distributed system requirements, it was nevertheless fruitful to look for the design assumptions that might fail in a networked environment or make recovery difficult. For example, a networked application might exhibit requirements for supporting disconnected operations by clients, and thus exhibit an architecture that supports a messaging communications model. A future study might explore how to leverage that type of architectural choice to support general survivability in the same way that this study leveraged survivability capabilities of the relational database infrastructure. In addition, the SNA method, artifacts, and lessons learned described in this case study report can be leveraged for survivability analysis of a variety of proposed and legacy DoD systems in diverse domains. These domains include distributed and networked command and control, integrated logistics, mission-specific, and real-time systems.

# 5 Acknowledgements

The authors are pleased to acknowledge the contributions of David Fisher and Howard Lipson of the CERT/CC to the definition of the SNA method.

The authors appreciate the support and cooperation of CarnegieWorks, Inc. in conducting this study. Cindi Carbine of CWI provided valuable insights into Vigilant system objectives, structure, and operational environment. The authors also acknowledge the team that developed the Sentinel subsystem prototype. The development was carried out as a studio project in the Master of Software Engineering program at Carnegie Mellon University. Team members included Melissa Cheok, Ryan Soulier, Grama Srinivasan, Shinji Yamato, and Isabel Yebra.

This report and the effort to produce it were sponsored by USAF Embedded Computer Resources Support Improvement Program (ESIP) and the U.S. Department of Defense.

24  CMU/SEI-98-TR-014

| | | |
|---|---|---|
| **REPORT DOCUMENTATION PAGE** | | *Form Approved OMB No. 0704-0188* |

Public reporting burden for this collection of information is estimated to average 1 hour per response, including the time for reviewing instructions, searching existing data sources, gathering and maintaining the data needed, and completing and reviewing the collection of information. Send comments regarding this burden estimate or any other aspect of this collection of information, including suggestions for reducing this burden, to Washington Headquarters Services, Directorate for information Operations and Reports, 1215 Jefferson Davis Highway, Suite 1204, Arlington, VA 22202-4302, and to the Office of Management and Budget, Paperwork Reduction Project (0704-0188), Washington, DC 20503.

| 1. **AGENCY USE ONLY** (LEAVE BLANK) | 2. **REPORT DATE** September 1998 | 3. **REPORT TYPE AND DATES COVERED** Final |
|---|---|---|
| 4. **TITLE AND SUBTITLE** A Case Study in Survivable Network System Analysis | | 5. **FUNDING NUMBERS** C — F19628-95-C-0003 |
| 6. author(s) R.J. Ellison; R.C. Linger; T. Longstaff; N.R. Mead | | |
| 7. **PERFORMING ORGANIZATION NAME(S) AND ADDRESS(ES)** Software Engineering Institute Carnegie Mellon University Pittsburgh, PA 15213 | | 8. **PERFORMING ORGANIZATION REPORT NUMBER** CMU/SEI-98-TR-014 |
| 9. **SPONSORING/MONITORING AGENCY NAME(S) AND ADDRESS(ES)** HQ ESC/DIB 5 Eglin Street Hanscom AFB, MA 01731-2116 | | 10. **SPONSORING/MONITORING AGENCY REPORT NUMBER** ESC-TR-98-014 |
| 11. **SUPPLEMENTARY NOTES** | | |
| 12.A **DISTRIBUTION/AVAILABILITY STATEMENT** Unclassified/Unlimited, DTIC, NTIS | | 12.B **DISTRIBUTION CODE** |

13. **ABSTRACT** (MAXIMUM 200 WORDS)

This paper presents a method for analyzing the survivability of distributed network systems and an example of its application. Survivability is the capability of a system to fulfill its mission, in a timely manner, in the presence of attacks, failures, or accidents. Survivability requires capabilities for intrusion resistance, recognition, and recovery. The Survivable Network Analysis (SNA) method builds on the Information Security Evaluation previously developed by permitting assessment of survivability strategies at the architecture level. Steps in the SNA method include system mission and architecture definition, essential capability definition, compromisable capability definition, and survivability analysis of architectural softspots that are both essential and compromisable. Intrusion scenarios play a key role in the method. SNA results are summarized in a Survivability Map which links recommended survivability strategies for resistance, recognition, and recovery to the system architecture and requirements. This case study summarizes the application and results of applying the SNA method to a subsystem of a large-scale, distributed healthcare system. The study recommended specific modifications to the subsystem architecture to support survivability objectives. Positive client response to study recommendations suggests that the method can provide significant added value for ensuring survivability of system operations. As a result of this case study, the SNA method, artifacts, and lessons learned will be available to apply architectural analysis for survivability to proposed and legacy DoD distributed systems.

| 14. **SUBJECT TERMS** architecture analysis, essential services, intrusion scenarios, network systems, survivability, Survivability Map | | | 15. **NUMBER OF PAGES** 24 pp. |
|---|---|---|---|
| | | | 16. **PRICE CODE** |
| 17. **SECURITY CLASSIFICATION OF REPORT** UNCLASSIFIED | 18. **SECURITY CLASSIFICATION OF THIS PAGE** UNCLASSIFIED | 19. **SECURITY CLASSIFICATION OF ABSTRACT** UNCLASSIFIED | 20. **LIMITATION OF ABSTRACT** UL |

NSN 7540-01-280-5500

Standard Form 298 (Rev. 2-89)
Prescribed by ANSI Std. Z39-18
298-102